\documentclass[a4paper,10pt]{article}
\usepackage[utf8x]{inputenc}

\title{General Formalism For the BRST Symmetry}
\author{Suhail Ahmad\\ Departmet of Physics, \\
Jamia Millia Islamia University,  New Delhi, India}

\begin{document}

\maketitle

\begin{abstract}
In  this paper we  will discuss Faddeev-Popov method for field 
theories with a gauge symmetry  in an abstract way.
We will then develope a general formalism for dealing with the 
BRST symmetry. This formalism will make it possible to
analyse the BRST symmetry for any theory. 
\end{abstract}

\section{Introduction}
It is not possible to 
directly quantize a field theory with gauge symmetry. In order 
to quantize such theories, we need to only sum over the physical 
field configurations and not the pure gauge ones.
This can be achieved by Faddeev-Popov method \cite{qp1}-\cite{qp2}. 
So,  we will discuss Faddeev-Popov method for field 
theories with a gauge symmetry  in an abstract way. This 
method gives rise to Faddeev-Popov ghosts. 
Ghosts occur in higher derivative theories  \cite{3a}-\cite{4a}
and gauge theories \cite{cam}-\cite{cam1}. A way to deal with these ghosts 
in gauge theory is called the BRST symmetry \cite{1a}-\cite{2a}. 
Recently,  BRST symmetry has been
 studied in gravity and M-Theory \cite{7a}-\cite{8a}.
In  BRST formalism  the sum of the classical 
Lagrangian, the gauge fixing term and the ghost term  is 
invariant the BRST transformations. This can be used to
remove all the negative norm states associated with the  
Faddeev-Popov ghosts. 
The BRST transformation of the original fields is there 
gauge transformation with the gauge parameter replaced by a 
ghost field.  The BRST transformation of the ghosts is give by 
the anti-commutator of those ghost fields.  The anti-ghosts 
transformation in auxiliary fields under BRST transformation and the 
BRST transformations of these auxiliary fields vanishes.  These BRST 
transformations are nilpotent.  This property of the BRST transformations 
is used to project out the physical sub-space of the gauge theory.
 This is done by defining the physical states as those states which 
are annihilated by the conserved charge generated by the invariance of 
the total Lagrangian under these BRST transformations.  
In this paper we will analyse   a gauge theory with a very 
general type of gauge transformations. All theories with gauge 
symmetry from Yang-Mills theory to gravity can be analysed as 
particular cases of this general gauge theory. Thus, our formalism 
is a very general formalism and we can see actually how the gauge
 fixing and ghost terms are generated for any gauge theory. We will 
also analyse how the BRST formalism works in general for 
any theory with gauge symmetry.

\section{Faddeev-Popov Method}

Let $\mathcal{A}^i  $ denote the field we are considering.
Now the classical action  $S$ of a theory is 
invariant under gauge transformations. Here $'i'$ denotes spacetime as
well as gauge indices.
Suppose the gauge transformations are given by
\begin{equation}
 \mathcal{A}^i  \to \mathcal{A}^i  + g^{i}_{jk}\Lambda^j\mathcal{A}^k,
\end{equation}
where $g^{i}_{jk}\Lambda^j\mathcal{A}^k$ is a general functional 
 of the infinitesimal parameter $\Lambda^i$. 
From now on we will supress the indices $'i'$.
Thus we have 
\begin{equation}
 \delta S_c =0, 
\end{equation}
where 
\begin{equation}
  \delta S_c = S_c [ \mathcal{A}  + g(\mathcal{A}, \Lambda) ] 
- S_c[ \mathcal{A}  ].
\end{equation}
This will lead to over counting and the divergence of the functional 
integral
\begin{equation}
 Z = \int D \mathcal{A}  D \Lambda 
 \exp i S_c[\mathcal{A}  ]
\end{equation}
where
\begin{equation}
Z \to \infty
\end{equation}
To fix this problem we want to restrict this path integral to
$F[\mathcal{A}  ] = 0$. 
This condition is called a  gauge fixing condition. 
 This is
achieved by inserting $\delta(F[\mathcal{A}  ])$ in 
the functional integral. First we note that
\begin{equation}
 Z = \int D \mathcal{A}  D \Lambda \delta(F[\mathcal{A}  '] )
 \det \left[\frac{\delta F[\mathcal{A}  ']}{\delta \Lambda} \right] 
\exp i S_c [\mathcal{A}  ].
\end{equation}
Here we have used 
\begin{equation}
 1 = \int D\Lambda \delta(F[\mathcal{A}  ']) 
\det \left[\frac{\delta F[\mathcal{A}  ']}{\delta \Lambda} \right].
\end{equation}
Then we define a function $F'[\mathcal{A}  ]$ as
\begin{equation}
 F'[\mathcal{A}  ] = F[\mathcal{A}  ] - \phi.
\end{equation}
Here $\phi(x)$ is a scalar function.
As $\phi$ does not depend on $\Lambda$, so we have
\begin{equation}
 \det \left[\frac{\delta F[\mathcal{A}  ]}{\delta \Lambda} \right] 
 =\det \left[\frac{\delta F'[\mathcal{A}  ]}{\delta \Lambda} \right].
\end{equation}
Now we can write
 \begin{equation}
 Z = \int D \mathcal{A}  D \Lambda \delta(F'[\mathcal{A}  '] )
 \det \left[\frac{\delta F'[\mathcal{A}  ']}{\delta \Lambda} \right] 
\exp i S_c [\mathcal{A}  ].
\end{equation}
Now we change the variables from $\mathcal{A}  $ to $\mathcal{A}  '$
 and as this is a simple shift, we have $D\mathcal{A}  = 
D\mathcal{A}  '$ and $S_c[\mathcal{A}  ] = S_c [\mathcal{A}  ']$. 
Now as $\mathcal{A}  '$ is a dummy variable we can 
rename it back to $\mathcal{A}  $ and obtain
\begin{equation}
 Z = \int D \mathcal{A}  D \Lambda \delta(F'[\mathcal{A}  ] ) 
\det \left[\frac{\delta F'[\mathcal{A}  ]}{\delta \Lambda} \right]
 \exp i S_c [\mathcal{A}  ].
\end{equation}
Now if $c$ and $\overline{c}$ are anticommuting fields then
\begin{equation}
  \det \left[\frac{\delta F'[\mathcal{A}  ]}{\delta \Lambda} \right] = 
 \int D c D\overline{c} \exp \left( -i \int d^4 x 
\sqrt{-g} \overline{c} \mathcal{L} c \right).
\end{equation}
Now as this holds for any $\phi(x)$, it will also hold for a 
normalized linear combination of $\phi(x)$ involving different 
$\phi(x)$. Now we integrate over all $\phi (x)$  as follows
\begin{equation}
  \int D\phi \exp \left( \frac{-1}{2\alpha} \int d^4 x \sqrt{-g}
 \phi^2 (x)\right) Z = N \int D \mathcal{A}  D\overline{c}
 D c  \exp i S_t,
\end{equation}
where $N$ is the normalization constant and $S_t$ is given by
\begin{equation}
 S_t = S_c + S_g + S_{gh}.
\end{equation}
Here $S_c$ is the original classical action,
 $S_g$ is the gauge fixing term 
and $S_{gh}$ is the ghost action.

\section{General Formalism}
We will now discuss the general formalism for BRST in an abstract way. 
To do so we write the gauge 
fixing term  by adding an auxiliary field $B$.
The gauge fixing Lagrangian with an auxiliary field $B$ is written as
\begin{equation}
 S_{g} = \int d^4 x \sqrt{-g}[- B F[\mathcal{A}  ] + \frac{\alpha}{2} B^2].
\end{equation}
$B$ does not contain any derivatives and the functional integral
 over $B$ can be done by completing the square and this way we will 
recover the gauge fixing term obtained by Faddeev-Popov method.
Now if we take the gauge transformation of the gauge fixing condition, 
 \begin{equation}
  \delta F[\mathcal{A}  ] = \mathcal{G}[\mathcal{A}  ],
 \end{equation}
 and replace $\mathcal{A}  $ by the ghosts $c$ to get $\mathcal{G}[c]$.
Then the ghost action is given by
 \begin{equation}
 S_{gh} = -i\int d^4 x \sqrt{-g}[ \overline{c} \mathcal{G}[c]].
 \end{equation}
The   action $S_t$ is invariant
 under a  symmetry call the BRST symmetry.
The BRST transformations are give by 
\begin{eqnarray}
  s   \mathcal{A}^i  &=&  i   g^{i}_{jk}c^j\mathcal{A}^k,\nonumber \\
 s   B^i &=& 0,\nonumber \\
 s   \overline{c}^i &=& B^i,\nonumber \\
 s   c^i &=& \frac{-i}{2}  f^{i}_{jk} c^i c^k.
\end{eqnarray}
Here the BRST transformation of $\mathcal{A}^i  $ is obtained by
 replacing the infinitesimal
 parameter $\Lambda^i$ in the gauge transformations by the
 ghost field $c^i$ and the BRST
 transformation of $c^i$ is obtained by taking 
the commutator of two gauge transformations and then replacing 
all the infinitesimal parameters by the $c^i$.
Thus the function $f^{i}_{jk}$, which is usually a constant, 
 is obtained by taking the 
commutator of the variation and then replacing $\Lambda^i$ by $c^i$.
The BRST transformation of $B^i$ vanishes and
 BRST transformation of $\overline c^i$ is $B^i$. 
These BRST transformations are nilpotent
\begin{eqnarray}
 s^2 \mathcal{A}^i &=& 0, \nonumber \\
 s^2 c^i &=& 0, \nonumber \\
 s^2 \overline c^i &=& 0, \nonumber \\
 s^2 B^i&=& 0.
\end{eqnarray}
This niloptency is important is isolating 
the physical states of the theory.
\section{Physical States}
Now we will discuss  general property of BRST charge.
It is known that there is a conserved charge called Noether's charge
 corresponding to each symmetry under which the action is invariant. 
The effective action which is formed by the sum of the original action, 
the gauge fixing term and the ghost term is invariant under the BRST
 transformation.  The Noether's charge corresponding to BRST 
transformation is the BRST charge $Q$
\begin{equation}
 Q = \int d^4x J^0,
\end{equation}
where 
\begin{equation}
 J^0 = \frac{\partial \mathcal{L}_t}{\partial \partial_0\mathcal{A}  } s \mathcal{A}  +
\frac{\partial \mathcal{L}_t}{\partial \partial_0c} s c +
\frac{\partial \mathcal{L}_t}{\partial \partial_0\overline c} s \overline c +
\frac{\partial \mathcal{L}_t}{\partial\partial_0 B} s B,
\end{equation}
and
\begin{equation}
 S_t = \int d^4 x \sqrt{-g} \mathcal{L}_t.
\end{equation}
It is nilpotent as its action 
on any field $|\mathcal{A}  \rangle$ twice vanishes.
\begin{equation}
 Q^2 |\mathcal{A}  \rangle = 0.
\end{equation}
Physical states $|P\rangle$  are  annihilated by $Q$
\begin{equation}
 Q |P\rangle = 0.
\end{equation}
Now physical states can be divided into two types; $| Pt\rangle$
 which are those physical state  that are obtained from the action
 of $Q$ on unphysical states $| UP\rangle $
\begin{equation}
| Pt\rangle = Q | UP\rangle,
\end{equation}
 and $| Pnt\rangle$ which are those physical states that are not obtained 
from the action of $Q$ on any state
\begin{equation}
| Pnt\rangle \neq Q | UP\rangle.
\end{equation}
It is obvious that
 any state that can be represented as the action of the BRST
 charge on any other state is a physical state, as it will be 
annihilated due to nilpotency of $Q$
\begin{equation}
 Q|Pt\rangle = Q^2 | UP\rangle= 0.
\end{equation}
All such states are infact orthogonal to all physical states including 
themselves
\begin{equation}
 \langle P| Pt\rangle =  \langle P| Q |UP\rangle = 0.
\end{equation}
All physical amplitudes involving such null states vanish. 
Two physical states that differ from each other by a null state
 will be indistinguishable,
 \begin{equation}
  |P'\rangle = |P\rangle + Q | \mathcal{A}  \rangle.
 \end{equation}
Thus  the revelant physical states in the theory 
are those are those physical states that are not obtained 
from the action of $Q$ on any other state i.e., $| Pnt\rangle$.
So we can identify the physical Hilbert space as a set of 
equivalence classes. This is how we factor out the physical 
state from the total Hilbert space of states. It is interesting 
to note that the nilpotency of these BRST transformations was 
crucial for isolating the physical states of the theory. 
If this nilpotency was broken then it would not be possible to isolate
the physical states of the theory. 

\section{Conclusion}
In this paper we have analysed a general formalism for the BRST symmetry.
The nilpotency of these BRST transformation can be used to 
isolate the physical state of the theory. 
This formalism can be used to analyse the BRST symmetry for gravity
and M-theory. 
It will be intresting to analyse the anti-BRST symmetry also in this 
general way. It will also be intresting to perform this analyses in
curved spacetime. To work out the BRST transformations in 
anti-de Sitter will be trivial as there are no IR divergences for 
the ghosts in anti-de Sitter spacetime \cite{air}-\cite{airq}. 
However, it will be difficult to
do it in de Sitter spacetime as there are 
IR divergences for 
the ghosts in anti-de Sitter spacetime \cite{ir}-\cite{ir1}.
It will also be intresting to analyse 
the third quantization \cite{zp}-\cite{zq}
of gravity using this BRST charge in analogy with string field theory
\cite{sft}-\cite{sft1}.

\end{document}